\documentclass[a4paper,12pt]{article}

\usepackage{amsmath}
\usepackage{amssymb}
\usepackage[dvips]{graphicx}
\usepackage[dvips]{psfrag}

\makeatletter
\@addtoreset{equation}{section}
\renewcommand{\theequation}{\thesection.\@arabic\c@equation}
\makeatother

\newcommand{\ssum}[2]{\mathop{\overset{#2}{\underset{#1}{\textstyle\sum}}}}

\newcommand{\tr}{\mathop{\mathrm{tr}}}
\newcommand{\e}{\mathrm{e}}
\newcommand{\rmd}{\mathrm d}

\newcommand{\Repa}{\mathop{\mathrm{Re}}}
\newcommand{\Impa}{\mathop{\mathrm{Im}}}
\newcommand{\Deriv}[2]{\frac{\rmd #1}{\rmd #2}}
\newcommand{\deriv}[2]{\frac{\partial #1}{\partial #2}}

\newcommand{\bracket}[1]{\left\langle #1\right\rangle}

\newcommand{\Const}{\text{const.}}
\newcommand{\eff}{\text{eff}}

\begin{document}

\titlepage

\vspace*{-15mm}   
\baselineskip 10pt   
\begin{flushright}   
\begin{tabular}{l}   
{December 2005} \\ 
{KUNS-2003}\\
{hep-th/0512176}   
\end{tabular}   
\end{flushright}   
\baselineskip 24pt   
\vglue 10mm   

\begin{center}
{\Large\bf
 Nonperturbative Effect \\
 in $c=1$ Noncritical String Theory \\
 and Penner Model
}

\vspace{8mm}   

\baselineskip 18pt   

\renewcommand{\thefootnote}{\fnsymbol{footnote}}

Yoshinori Matsuo \footnote[2]{E-mail: ymatsuo@gauge.scphys.kyoto-u.ac.jp}

\renewcommand{\thefootnote}{\arabic{footnote}}
 
\vspace{5mm}   

{\it  
 Department of Physics, Kyoto University, 
 Kyoto 606-8502, Japan 
}
  
\vspace{10mm}   

\end{center}

\begin{abstract}
We derive the effect of instantons in the Penner model. 
It is known that the free energies of the Penner model and 
the $c=1$ noncritical string at self-dual radius 
agree in a suitable double scaling limit. 
On the other hand, the instanton in the matrix model describes 
a nonperturbative effect in the noncritical string theory. 
We study the correspondence between 
the instantons in the Penner model 
and the nonperturbative effect 
in $c=1$ noncritical string at self-dual radius. 
\end{abstract}

\baselineskip 18pt   

\newpage

\section{Introduction}\label{sec:intro}

After the discovery of the D-brane, 
which is the typical nonperturbative effect in the string theory, 
the study of the nonperturbative effect is of great interest. 
We can analyze the noncritical string theory, 
which is a simplified model of the string theory, 
nonperturbatively via the matrix model (for reviews, see 
\cite{DiFrancesco:1993nw, Ginsparg:1993is, Klebanov:1991qa}). 
The nonperturbative effect in noncritical string theory 
can be derived via the string equation. 
On the other hand, 
this effect appears as the instanton 
when we directly analyze the matrix model. 
The string equation cannot describe the whole of 
the nonperturbative effect. 
There is an ambiguity that corresponds to 
the initial condition of the string equation. 
Studying the matrix model directly, 
we can fix it \cite{Hanada:2004im}. 

This analysis is generalized to various models 
\cite{Kawai:2004pj, Sato:2004tz, Ishibashi:2005zf}. 
One of the most interesting models is 
the $c=1$ noncritical string theory 
because it describes 
the dynamics of strings in two dimensional space-time. 
In this paper, 
we consider the Penner model, which 
corresponds to the $c=1$ noncritical string theory 
with one direction of the space-time compactified on 
the circle with self-dual radius \cite{Distler:1990mt}. 

The Penner model is defined as a kind of the hermitian one-matrix model. 
Therefore, we can extend the analysis for the $c=0$ matrix model 
to our case with a small modification. 
And, it is easy to see that 
the relation between the nonperturbative effect 
for the $c=0$ matrix model 
and that in the Penner model. 
In this paper, 
we generalize the analysis of the instanton in the $c=0$ matrix model 
to the case of the Penner model. 
And we show that some of the nonperturbative effect 
in the $c=1$ noncritical string theory 
can be understood as the effects of instantons in the Penner model.

This paper is organized as follows. 
First, in section~\ref{sec:c=0}, 
we consider the instanton in the $c=0$ matrix model. 
We also give the interpretation which makes it possible 
to extend the analysis to the case of the Penner model. 
In section~\ref{sec:penner}, we summarize 
some of the known facts related to the correspondence 
between the Penner model and the $c=1$ noncritical string theory. 
In section~\ref{sec:inst}, we make the redefinition of 
the Penner model to introduce the instantons, 
and calculate the contribution from the instantons 
to the free energy. 
In section~\ref{sec:concl} is devoted for 
the conclusions and discussions. 
In appendices, we show the details of calculations 
and the behaviors of orthogonal polynomials in large $N$ limit.

\section{The instantons in $c=0$ noncritical string theory}\label{sec:c=0}

Before studying the effect of instantons 
for the $c=1$ noncritical string theory,
in this section, we review the results obtained in the case of 
the $c=0$ noncritical string theory \cite{Hanada:2004im}. 
We focus on the contribution of instantons 
to the free energy. 
In order to extend the instanton calculation to the $c=1$ case,
we also give the precise definition of instanton. 

The $c=0$ noncritical string theory can be 
described by the hermitian one-matrix model. 
Its partition function is given by 
\begin{equation}
 Z = \int\rmd\Phi \,\e^{-N\tr V(\Phi)},
\end{equation}
where $\Phi$ is an $N\times N$ hermitian matrix. 
By diagonalizing the matrix $\Phi$, this theory can be 
described by eigenvalues of the matrix $\Phi$, 
and the partition function can be expressed as
\begin{equation}
 Z = \int\prod_i\rmd\lambda_i\Delta^2(\lambda)\,\e^{-N\sum_iV(\lambda_i)}.
\end{equation}
Vandermonde determinant is defined as 
$\Delta(\lambda)=\prod_{i<j}(\lambda_i-\lambda_j)$.
Here, we concentrate on the $N$-th eigenvalue $\lambda_N$
and represent it $x$. 
Integrating out the other $N-1$ eigenvalues, 
we obtain the effective potential $V_\eff(x)$ of 
the $N$-th eigenvalue $x=\lambda_N$. 
The partition function can be expressed using 
the effective potential as 
\begin{equation}
 Z=\int\rmd x\,\e^{-NV_{\eff}(x)}
  =Z_{N-1}\int\rmd x\bracket{\det(x-\Phi)^2}_{N-1}\e^{-NV(x)},
  \label{EffectivePotential}
\end{equation}
where the subscript ``$N-1$'' indicate that 
the matrix $\Phi$ is replaced by 
an $(N-1) \times (N-1)$ matrix.
In the large $N$ limit, 
the system of an $N\times N$ matrix gives the 
same results as one of an $(N-1) \times (N-1)$ matrix. 
Hence, we can replace it by the expectation value of the standard 
$N\times N$ matrix 
\begin{equation}
 \bracket{\mathcal O} 
  = \frac{1}{Z}\int\rmd\Phi\ \mathcal O\,\e^{-N\tr V(\Phi)}.
\end{equation}

In the large $N$ limit, the effective potential $V_\eff(x)$ 
can be expressed using the resolvent 
$R(x)= \frac{1}{N}\bracket{\tr\frac{1}{x-\Phi}}$ as
\begin{align}
 V_\eff(x) &= V(x) -2\Repa\int^x\rmd x'\,R(x') \notag\\
  &=\Repa\int^x\rmd x'\sqrt{V^{\prime 2}(x')+4f(x')},
\end{align}
at the leading order of the $\frac{1}{N}$ expansion. 
Now we consider the model corresponding to 
the $c=0$ noncritical string theory. 
Taking the double scaling limit, we obtain
\begin{equation}
 V_\eff(x) =\Const\times\Repa\int^x\rmd x'\left(x'-\frac12\sqrt t\right)
  \sqrt{x'+\sqrt t} .
\end{equation}
The first derivative of the effective potential 
vanishes in two regions. 
One is the region where the resolvent has the cut. 
The effective potential takes a constant value in this region, 
and it is local minimum. 
The other is the point of $x=\frac{1}{2}\sqrt{t}$
where the effective potential has local maximum. 
The region of the cut gives the dominant contribution, 
which corresponds to the perturbative expansion. 
We call this region as ``dominant saddle point.''
And the point of the local maximum of the effective potential 
gives sub-dominant contribution, 
which corresponds to the nonperturbative effect. 
We call this point as ``sub-dominant saddle point,'' 
and this sub-dominant contribution, 
or the configuration in which one or more eigenvalues are 
at this ``sub-dominant saddle point,'' as ``instanton.''

Up to this point, we have restricted ourselves 
to the $N$-th eigenvalue. 
However, there are $N$ eigenvalues and all of them can be possibly 
at the ``sub-dominant saddle point.''  
Including these, we obtain
\begin{subequations}
 \begin{align}
 Z &= Z^{(\text{0-inst})}+Z^{(\text{1-inst})}+Z^{(\text{2-inst})}+\cdots , \\
 Z^{(\text{0-inst})} &= \int_{\text{inside the cut}}
 \prod_i\rmd\lambda_i\,
 \Delta^2(\lambda)\,\e^{-N\ssum{i}{}V(\lambda_i)}\notag\\
 &= Z^{(\text{0-inst})}_{N-1}\int_{\text{inside the cut}}\rmd x
 \bracket{\det(x-\Phi)^2}^{(\text{0-inst})}\e^{-NV(x)} , \\
 Z^{(\text{1-inst})} &= NZ^{(\text{0-inst})}_{N-1}
 \int_{\text{outside the cut}}\rmd x
 \bracket{\det(x-\Phi)^2}^{(\text{0-inst})}\e^{-NV(x)}.
 \end{align}
\end{subequations}
We have divided the integration region, namely, 
inside the cut and outside the cut. 
The ``dominant saddle point'' is identical to the region 
inside the cut, and the ``sub-dominant saddle point'' 
is included in the region outside the cut. 
The superscript ``0-inst'' indicates that 
all eigenvalues are inside the cut and 
``$n$-inst'' indicates that 
$n$ eigenvalues are outside the cut. 

By neglecting the interaction between the eigenvalues, 
the free energy becomes
\begin{subequations}
 \begin{align}
  e^{F} = Z &= Z^{(\text{0-inst})}
  \left(1+\frac{Z^{(\text{1-inst})}}{Z^{(\text{0-inst})}}+\cdots\right)
  = \e^{F^{(\text{0-inst})}+\delta F} , \\
  & F^{(\text{0-inst})} = -\frac{4}{15}t^\frac52 \\
  &\delta F = \frac{Z^{(\text{1-inst})}}{Z^{(\text{0-inst})}}
  =\frac{i}{16\cdot3^\frac34\sqrt\pi t^\frac58}
  \exp\left[-\frac{8\sqrt3}{5}t^\frac54\right].
 \end{align}
\end{subequations}

\begin{figure}[h]
 \begin{center}
  \includegraphics*[scale=0.6]{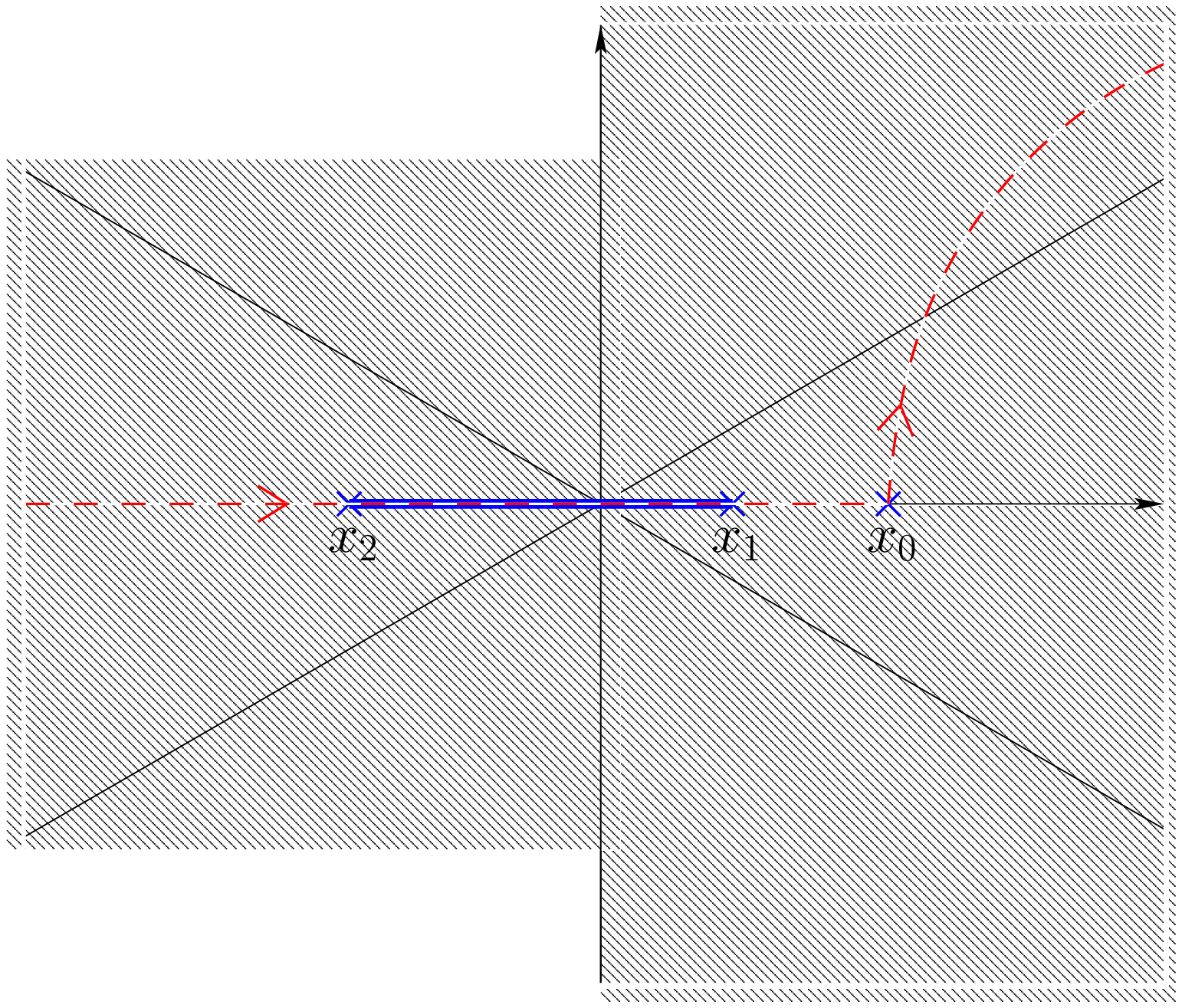}
 \end{center}
 \caption{The cut of the resolvent (the bold line) and 
 the contour of the integration (the dashed line) 
 corresponding to the $c=0$ noncritical string theory. 
 Here, we consider a cubic potential for instance. 
 We should put the beginning and the end of the contour at 
 $|z|\to\infty$ in the shaded region.}
 \label{contour_of_c=0}
\end{figure}

Now, we consider why the nonperturbative correction of 
the free energy becomes imaginary. 
The partition function of the matrix model is 
defined as an integration of a real function. 
By this definition, the free energy of 
this model should be real. 
To obtain the imaginary number, 
some change of the definition is needed. 
The imaginary part of the free energy reflects 
the instability of the model. 
In the $c=0$ matrix model, 
all the eigenvalues distribute around one minimum of the potential. 
This minimum corresponds to the ``dominant saddle point.'' 
However, the potential of the $c=0$ matrix model 
has some local minima other than this, 
or it is unbounded from below. 
To prevent the eigenvalues from distributing 
around these unwanted minima, or the unbounded region, 
we should deform the contour of the integration. 
We integrate the effective potential over $x$ 
along the real axis around the minimum 
corresponding to the ``dominant saddle point,'' 
and the contour turns at right angle after reaching 
the ``sub-dominant saddle point.''(Fig.~\ref{contour_of_c=0}) 
Then, we have the integration orthogonal to 
the real axis at the ``sub-dominant saddle.'' 
It gives an imaginary number. 
This modification of the curve of the integration 
is needed for all eigenvalues, 
and chosen not to include the unwanted vacua. 
Because we can define the partition function 
by the integration of the analytic function, 
the partition function depends only on 
the beginning and the end of the contour.

For example, we take a cubic potential 
$V(x) = \frac{1}{2}x^2-\frac{g}{3}x^3$
In this case, the potential is unbounded from below. 
Then, we should define the integration 
so as to avoid this divergence. 
On the other hand, 
the partition function is defined as 
\begin{equation}
 Z = \int\prod_i\rmd\lambda_i\Delta^2(\lambda)\e^{-N\sum_iV(\lambda_i)}.
\end{equation}
If we define the contour of 
the integration to be along the real axis, 
the integrand is diverges at $x\to\infty$. 
Generally, we assume that 
the contribution from boundary terms of integration 
vanishes 
in the $c=0$ matrix model. 
It can be done by taking 
the beginning and the end of the contour 
to be $|\lambda|\to\infty$ in the shaded region 
in Fig.~\ref{contour_of_c=0} | 
$-\frac\pi 2<\arg\lambda<-\frac{1}{6}\pi$,
$\frac{1}{6}\pi<\arg\lambda<\frac \pi 2$ or
$\frac{5}{6}\pi<\arg\lambda<\frac{7}{6}\pi$.
Now, we put the beginning of the contour 
in the region 
$|\lambda|\to\infty$ and 
$\frac{5}{6}\pi<\arg\lambda<\frac{7}{6}\pi$, 
and the end of the contour in 
$|\lambda|\to\infty$ and 
$\frac{1}{6}\pi<\arg\lambda<\frac\pi 2$.
Then, we obtain the wanted integration: 
the dominant contribution comes from the cut. 
It correspond to the $c=0$ noncritical string theory. 
If we put the beginning and the end of the contour 
in the region 
$-\pi<\arg\lambda<-\frac{1}{6}\pi$ and
$\frac{1}{6}\pi<\arg\lambda<\pi$, 
integration does not include the contribution from the cut. 
This definition of the integration 
does not correspond to the $c=0$ noncritical string theory. 
Thus, the choice of the beginning and the end of the contour 
determines which vacua are included in the theory. 

We take the contour corresponding to 
the $c=0$ noncritical string theory. 
The contributions from the perturbative expansion are 
divergent series, 
and the instantons give 
corrections to this series. 
Therefore, it isn't well-defined. 
But If we restrict ourselves on the imaginary part of 
the free energy, the situation becomes different. 
The contributions of a perturbative expansion are 
real because they come from the integration 
inside the cut. 
Hence, the dominant contribution of the imaginary part 
comes from the instanton. 
Therefore, the imaginary part of 
the nonperturbative effect is well-defined. 
In this way, we have obtained 
the imaginary part of the nonperturbative effect.

\section{The Penner model and the $c=1$ noncritical string}\label{sec:penner}

In this section, we study the Penner model. 
The Penner model is defined as 
a hermitian one-matrix model
with a logarithmic potential. 
It is noted that in the suitable double scaling limit, 
the free energy of the Penner model 
is identical to that of 
the $c=1$ noncritical string theory 
with the target space compactified on the circle 
of self-dual radius $R=1$ \cite{Distler:1990mt, Chaudhuri:1991hv}.
Here, we review the relation 
between the Penner model and $c=1$, $R=1$ noncritical string theory. 

The Penner model is defined as a hermitian one-matrix model 
with the following potential: 
\begin{equation}
 V(x) = \frac{1}{g}\left(x- \log x\right).\label{PennerPotential}
\end{equation}
We can derive the free energy via 
the method of orthogonal polynomials. 
This method makes use of an infinite set of 
polynomials obeying the orthogonality condition: 
\begin{equation}
 (P_n(x),P_m(x)) = \int\rmd x\,P_n(x)P_m(x)\,\e^{-NV(x)} = h_n\delta_{nm}.
  \label{orth}
\end{equation}
The normalization of orthogonal polynomials is given by having 
leading term $P_n(x) = x^n + \cdots$.
The partition function of the hermitian one-matrix model 
is given by 
\begin{equation}
 Z = N!\prod_{n=0}^{N-1} h_n. 
\end{equation}
So, we should calculate $h_n$. 
Orthogonal polynomials satisfy the following two recursion relations,
\begin{subequations}\label{RecursionRelation}
 \begin{align}
  xP_n(x) &= X_{nm}P_m(x) = P_{n+1}(x) + s_n P_n(x) + r_n P_{n-1}(x), 
  \label{Recursion1}\\
  P_n'(x) &= \mathcal P_{nm}P_m(x) = \left[NV'(X)\right]_{nm}P_m(x).
  \label{Recursion2}
 \end{align}
\end{subequations}
Using these relations, we can easily see that 
$r_n = \frac{h_n}{h_{n-1}}$. 
For the case of the Penner model \eqref{PennerPotential}, 
these recursion relations are exactly soluble. 
Solutions of these relations are 
\begin{subequations}
 \begin{align}
  r_n &= g\xi(g\xi+1) \\
  s_n &= 1+2g\xi+\frac{g}{N}.
 \end{align}
\end{subequations}

The partition function of the model is given by $Z = N!\prod_n h_n$,
and the free energy is
\begin{equation}
 F = N\sum_n \left(1-\frac{n}{N}\right)\log r_n.
\end{equation}
To obtain the free energy identical to 
the $c=1$ and $R=1$ noncritical string theory, 
we take the following double scaling limit 
\begin{align}
 N(1+g) &= t ,& N&\to\infty ,  
\end{align}
and replace the summation over $n$ 
by integration with respect to $\xi = \frac{n}{N}$ 
via Euler-Maclaurin summation formula.
Then, we obtain 
\begin{equation}
 F = \frac{1}{2}t^2\log t
  -\frac{1}{12}\log t + \sum_{h=1}^\infty\frac{B_{2h}}{2h(2h-2)}t^{2-2h},
\end{equation}
up to regular terms.
Taking $t=i\mu$, we obtain
\begin{equation}
 F = -\frac{1}{2}\mu^2\log\mu
  -\frac{1}{12}\log\mu
  + \sum_{h=1}^\infty\frac{\left\lvert B_{2h}\right\rvert}{2h(2h-2)}\mu^{2-2h}.
\end{equation}
This expression is identical to the free energy of 
the $c=1$ noncritical string theory compactified on the circle 
with self-dual radius. 

To see the precise behavior of the Penner model, 
we consider the resolvent and the concrete expressions 
of orthogonal polynomials. 
First, we consider the resolvent. 
In the large $N$ limit, 
we can derive the resolvent 
$R(x) = \frac{1}{N}\bracket{\tr\frac{1}{x-\Phi}}$
from the loop equation, 
and we obtain 
\begin{align}
 R(x) &= \frac{1}{N}\bracket{\tr\frac{1}{x-\Phi}} \notag\\
 &= \frac{1}{2gx}
 \left( x-1 \pm \sqrt{x^2-2(1+2g)x +1 }\right).
\end{align}
Here, the sign of ``$\pm$'' is depends on the location of the cut. 
The position of the cut is 
identical to the support of the eigenvalue density. 
It is determined by the following function $G(x)$. 
\begin{equation}
 G(x) = \int_{x_1}^x\rmd x'\left(V'(x')-2R(x')\right)
\end{equation}
The support of the eigenvalue density 
is an arc connecting two branch points of 
the resolvent satisfying the following two conditions: 
the real part of $G(x)$ vanishes along this support and 
the real part of $G(x)$ is negative around this support 
\cite{Chaudhuri:1991hv, Ambjorn:1994bp}. 
Here, we put $x_1$ on one of branch points of the resolvent. 

The Penner model is defined with positive coupling constant $g>0$ 
in \eqref{PennerPotential}. 
We also restrict the integration to positive eigenvalues, 
from $\lambda=0$ to $\lambda\to + \infty$. 
Therefore there are contributions the boundaries 
at $\lambda=0$ and $\lambda\to+\infty$. 
These contributions vanish for $g>0$. 
However, to obtain the model corresponding to 
the $c=1$ noncritical string theory, 
$g$ should be negative. 
We calculate the free energy, or some expectation value, 
with positive $g$, 
and at the end, analytically continue 
the result to negative $g$. 

In this process, we use the fact 
that boundary terms of the integration vanish 
for positive $g$.  
However, under the analytic continuation to negative $g$, 
it is merely an assumption that boundary terms can be dropped. 
In fact, if we use the contour of the integration 
which is used in the case of positive $g$, 
boundary terms do not vanish for negative $g$. 
Therefore we should deform the contour to drop boundary terms. 

\begin{figure}[h]
 \begin{center}
  \includegraphics[scale=0.5]{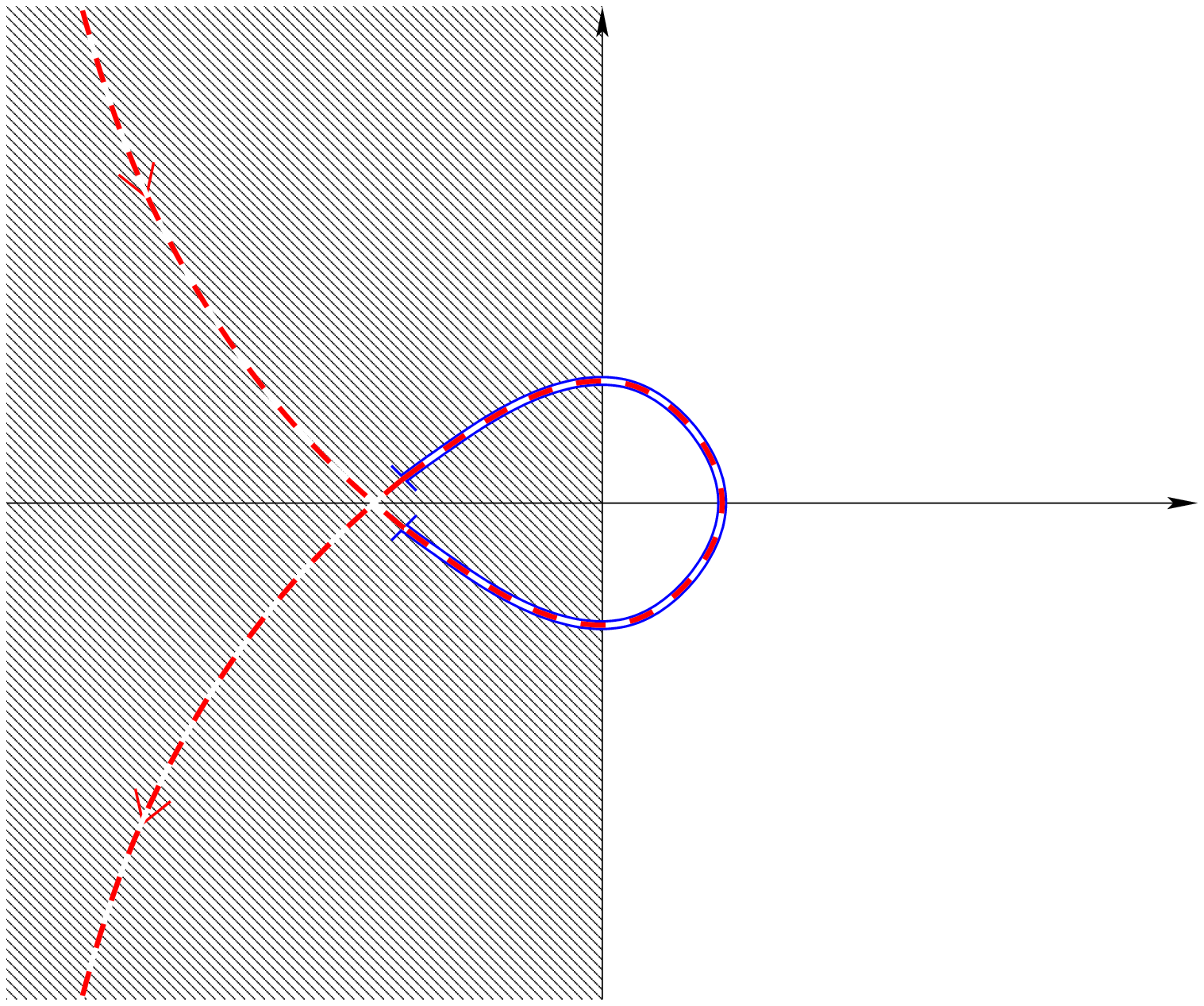}
  \caption{The contour of the integration (the dashed line) and 
  the cut of the resolvent (the bold line) 
  for the Penner model with negative $g$. 
  We should put both ends of the contour at $|\lambda|\to\infty$ 
  in the shaded region. 
  The support of the eigenvalue distribution is 
  along the contour, and no longer on the real axis.}
  \label{fig:contour_of_penner}
 \end{center}
\end{figure}

As is the case with the model corresponding to 
the $c=0$ noncritical string theory, 
we can determine the contour of the integration 
by specifying the position of its both ends. 
For the Penner model, boundary terms vanish 
when we put the boundary in the region where 
$|\lambda|\to\infty$ with real part of eigenvalues negative. 
Because the beginning and the end of the contour are 
located at the same point, 
the contour should go around the singularity of 
the logarithmic potential of the Penner model 
to obtain a nontrivial result. 
Then, the contour of the integration becomes as 
indicated in Fig.~\ref{fig:contour_of_penner}. 
The position of the cut changes depends on the choice of the contour. 
For positive $g$, the cut is located on the positive real axis 
and the contour can be taken to be along the cut. 
For negative $g$, both ends of the cut have the imaginary part, 
and the cut intersect the real axis on 
its positive part.(Fig.~\ref{fig:contour_of_penner}) 
The result of the integration doesn't change under 
the deformation of the contour 
which doesn't move the both ends and doesn't cross singularity. 
And the position of the cut has been determined 
when we have specified the position of the both ends of the contour. 
Then, we can take the contour to be along the cut. 

Next, we consider the concrete expression of 
orthogonal polynomials. 
We show only the result here, 
and the detailed calculations are given in 
Appendix~\ref{app:orth_poly}. 
In large $N$ limit, 
orthogonal polynomials can be written as 
\begin{align}
 \psi_n(x) &= \frac{1}{\sqrt{h_n}} P_n(x)\,\e^{-\frac{1}{2}NV(x)} \notag\\
  &= \left(\frac{(x-(1+2g\xi)+\sqrt{x^2-2(1+2g\xi)x+1})^2}
 {4g\xi(g\xi+1)(x^2-2(1+2g\xi)+1)}\right)^{\frac{1}{4}}
  \exp\left[-\frac{N\xi}{2}G_{\xi}(x)\right],
\end{align}
where, $G_\xi(x)$ is $G(x)$ with $g$ replaced by $g\xi$.

\section{Instantons in the Penner model}\label{sec:inst}

In this section, we consider the effect of the instanton 
in the Penner model. 
In the case of the $c=0$ matrix model, 
the model is a hermitian one-matrix model, 
and its dynamics can be reduced to that of the eigenvalues. 
The instanton is defined as the configuration 
in which one of the eigenvalues is located at 
the local maximum of the effective potential. 
The Penner model is also a hermitian one-matrix model. 
We can use the same definition of the instanton 
as that for the case of the $c=0$ theory. 

The effective potential of an eigenvalue of 
a hermitian one matrix model is defined as 
\begin{equation}
 Z=\int\rmd x\,\e^{-NV_{\eff}(x)}
  =Z_{N-1}\int\rmd x\bracket{\det(x-\Phi)^2}_{N-1}\e^{-NV(x)}.
  \tag{\ref{EffectivePotential}}
\end{equation}
The effective potential can be expressed 
in terms of orthogonal polynomials as 
\begin{equation}
  \e^{-NV(x)}\bracket{\det\left(x-\Phi\right)^2} 
  = h_N \sum_{n=0}^N\psi_n^2(x), 
  \label{EffectivePotential_OrthogonalPolynomial}
\end{equation}
where the orthonormal functions 
$\psi_n(x) = \frac{1}{\sqrt{h_n}} P_n(x)\,\e^{-\frac{1}{2}NV(x)}$ 
are build from the orthogonal polynomials. 
The ``dominant saddle'' of this effective potential 
gives the contribution to the partition function 
that corresponds to the perturbative expansion. 
The orthonormal functions can be approximated 
in the large $N$ limit as 
\begin{equation}
 \psi_n(x) =
  \exp\left[-\frac{N\xi} 2
       G_{\xi}(x)\right]. \label{LargeNOrthFunc}
\end{equation}
Therefore, the ``dominant saddle'' of $\psi_n^2$ 
is located at the position of the cut of the resolvent 
with $g$ replaced by $g\xi$. 
For the model corresponding to the $c=0$ noncritical string theory, 
the position of the cut of $\psi_n(x)$ for $n < N$ is 
included in the position of the cut of $\psi_N(x)$. 
Therefore, we can divide the integration into two parts: 
the inside the cut of and outside of the cut of $\psi_N(x)$. 
On the other hand, for Penner model, 
the position of the cut is varies with $n$. 
But, the contour of the integration can be deformed 
because of the analyticity of $\psi_n(x)$. 
So, we first decompose the integration into 
that of the individual $\psi_n(x)^2$, 
and then deform the contour to be along the cut. 
In this way, we can pick up the contribution from the 
``dominant saddle'' of all of $\psi_n(x)^2$. 
These contributions reproduce 
the perturbative expansion of the $c=1$ noncritical string theory. 

\begin{figure}[h]
 \begin{center}
  \includegraphics*[scale=0.6]{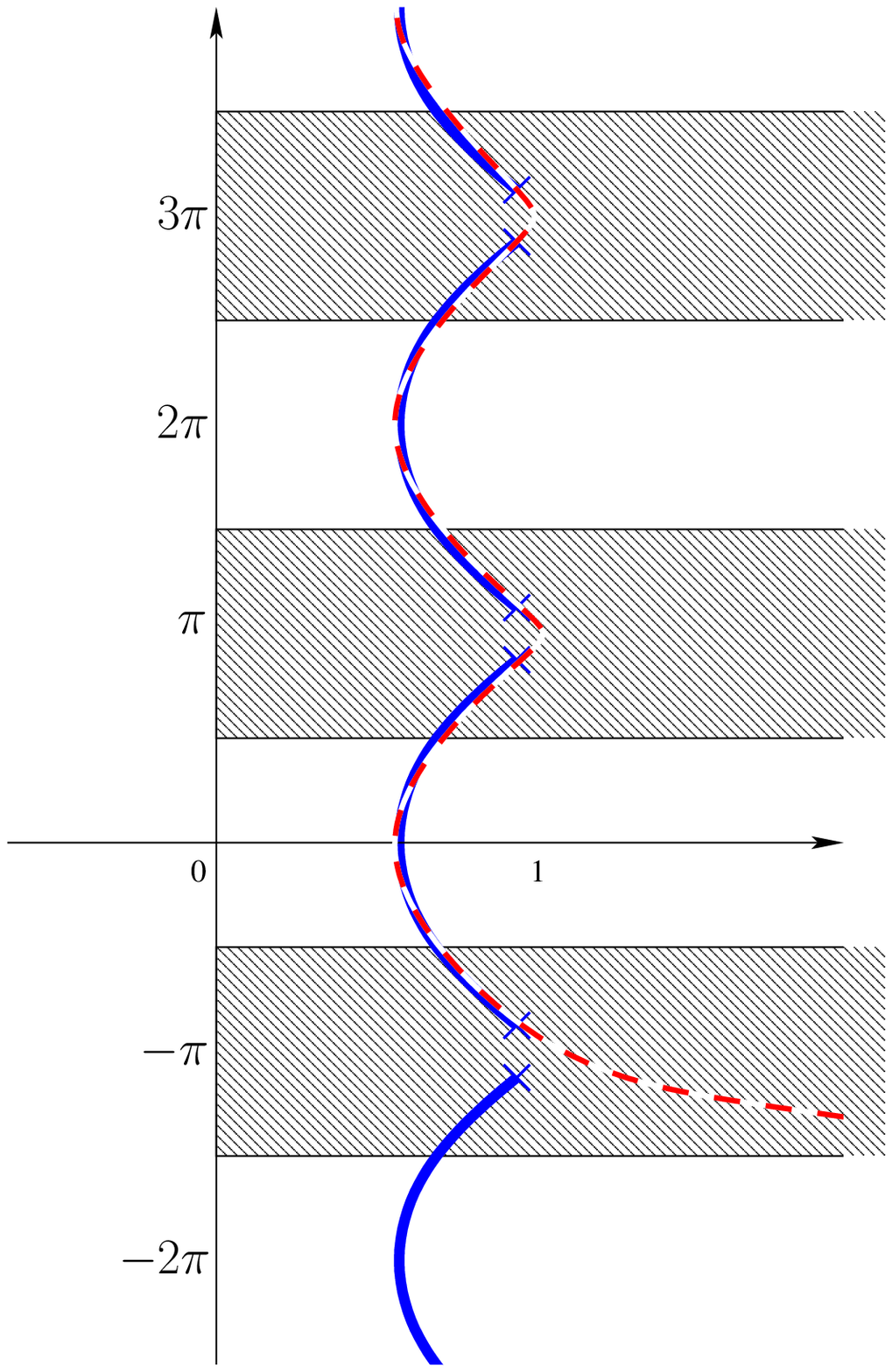}
  \caption{The contour of the integration which we consider 
  (the dashed line) and 
  cuts of the resolvent (the bold line) in polar coordinate. 
  There are infinite number of the cuts 
  and these cuts give the different contribution 
  to the partition function each other. 
  We take the beginning of the contour in the region 
  $-\frac{3}{2}\pi<\arg x<-\frac{1}{2}\pi$ and $|x|\to\infty$, 
  and the end of the contour in the region $\arg x\to \infty$. 
  The integration picks up all cuts in the region 
  $(2n-1)\pi<\arg x<(2n+1)\pi$ for $n\geq 0$. 
  The cut with $n=0$ gives the contribution corresponding to 
  the perturbative series of the $c=1$ noncritical string theory. 
  The cuts with $n\geq 1$ correspond to 
  the nonperturbative effect. 
  }\label{contour_of_c=1}
 \end{center}
\end{figure}
There is no local maximum in the effective potential 
for the Penner model except for what comes from 
the cut of the resolvent. 
However, the effective potential is multivalued on 
the complex plane of the $N$-th eigenvalue $x$, 
due to the logarithmic term of the potential. 
So, we use a polar coordinate on which 
the effective potential becomes 
single-valued.\footnote{
In the large $N$ limit, 
the effective potential is expressed in terms of the resolvent. 
Because the resolvent has the cut, 
the effective potential seems to be multivalued. 
However, this multivaluedness is due to the approximation 
that is justified only in the large $N$ limit. 
$\exp\left[-N\left(V_\eff(x)-V(x)\right)\right]$ 
can be expressed in terms of the polynomials 
\eqref{EffectivePotential_OrthogonalPolynomial}, 
and therefore the effective potential doesn't have 
the multivaluedness except 
what comes from the potential of the model. 
} 
When we extend the range of $\theta=\arg x$ 
to $-\infty<\theta<+\infty$, 
there is a cut for each additional $2\pi$ of $\theta$. 
These infinite number of the cuts can be counted as 
the ``saddle point'' of the effective potential. 
For the Penner model, 
we usually assume that boundary terms of 
the integration with respect to the eigenvalues vanish. 
In the case for negative $g$, 
each end of the contour should be in the region in which 
$|x|\to\infty$ and $\Repa x < 0$. 
When, the range of $\theta$ is extended to $(-\infty,+\infty)$, 
each end of contour can be at any of 
$|x|\to\infty$ in shaded region of Fig~\ref{contour_of_c=1}
($\left(2n+\tfrac{1}{2}\right)\pi<\theta<\left(2n+\tfrac{3}{2}\right)\pi$ 
with $n=\cdots,-2,-1.0,1,2,\cdots$). 
To make the model correspond to the $c=1$ noncritical string theory, 
we should add a small imaginary part to $g$, 
as we have seen in previous section. 
Then, boundary terms vanish at $\theta\to\infty$. 
We put the beginning of the contour at $|x|\to\infty$ 
in one of the shaded region of Fig~\ref{contour_of_c=1}, 
and the end of the contour at $\theta\to\infty$ 
(Fig~\ref{contour_of_c=1}). 
This contour picks up the cut which lies in the region where 
$\theta$ is greater than the beginning of the contour. 
The dominant contribution comes from the cut 
with the lowest value of the effective potential, 
which is located at the smallest $\theta$ of all the cuts 
along the contour, 
and the other cuts give sub-dominant contributions. 

Now, we calculate the dominant contribution and 
nonperturbative corrections to the free energy of the Penner model. 
First, integrating the effective potential for the $N$-th eigenvalue, 
we calculate the nonperturbative correction from the configuration
in which one eigenvalue is instanton. 
The effective potential can be written as 
\begin{equation}
 \e^{-NV_\eff(x)} = \sum_{n=0}^N \psi_n^2(x), 
\end{equation}
where $\prod_n h_n$ is omitted because it does not depend on $x$. 
In the large $N$ limit, 
the orthonormal function $\psi_n(x)$ 
is constant on the cut up to sub-leading order, 
because $\psi_n(x)$ can be expressed as 
\eqref{LargeNOrthFunc}. 
We call the cut located in the region $(2k-1)\pi<\theta<(2k+1)$ 
$k$-th cut and denote it by $\mathcal C_k$. 
We define the value of $G_\xi(x)$ on these cuts as 
\begin{align}
 \xi G_\xi(x_k) &\equiv \bar G_k(\xi) = \Const ,&
 &\text{for }x_k\in\mathcal C_k .
\end{align}
Because the orthogonal polynomials $P_n(x)$ are single valued 
on the complex plane of $x$, 
the orthonormal functions $\psi_n(x)$ become multivalued via 
the potential of the Penner model $V(x)$. 
Therefore, the next to leading order corrections in $\frac{1}{N}$ 
of $\psi_n(x)$ is single valued. 
In fact, we can see that 
the $\mathcal O(N^0)$ contribution 
of the orthogonal polynomials are single valued, 
via their concrete expression in the large $N$ limit (see appendix). 
Then, integrating the effective potential, 
we obtain, 
\begin{align}
 \int\rmd x\,\e^{-NV_\eff(x)} &= \int\rmd x\sum_{n=0}^N\psi_n^2(x) \notag\\
 &= \sum_{n=0}^N\sum_{k=0}^{\infty}
 \int_{\mathcal C_0}\rmd x\,\psi_n^2(x)
 &= \sum_{n=0}^N\sum_{k=0}^{\infty}
 \e^{-N\left[\bar G_k(\frac{n}{N})-\bar G_0(\frac{n}{N})\right]}
 \int_{\mathcal C_0}\rmd x\,\psi_n^2(x) . 
\end{align}
Here, we define the normalization of the orthonormal functions 
in such a way that $\bar G_0=0$, namely, 
\begin{equation}
 \int_{\mathcal C_0}\rmd x\psi_n^2(x)=1. 
\end{equation}

To obtain the result corresponding to the 
$c=1$ noncritical string theory, 
we should take the suitable double scaling limit. 
We perform the rest of the calculation in this limit. 
We take the following limit, 
\begin{align}
  g&=-1+a^2t, &
 x&=-1+az, &
 \xi & = 1-a^2\eta, &
 N&= a^{-2} .
\end{align}
In this limit, the difference of $G_\xi(x)$'s 
between two neighboring cuts 
becomes, 
\begin{align}
 \bar G_{k+1}(\xi)-\bar G_k(\xi) &= \int\rmd z\sqrt{z^2+4(t+\eta)} \notag\\
 &= -2\pi i(t+\eta) . \label{G-G}
\end{align}
For the Penner model to correspond to 
the $c=1$ noncritical string theory, 
we should take $t=i\mu$, 
and it becomes $\bar G_{k+1}(\xi)>\bar G_k(\xi)$. 
We choose the contour in such the way that 
we pick up the $k$-th cut from $k=0$ to $k\to\infty$ 
in the integration. 
The dominant contribution comes from 
the $0$-th cut, 
because the $\bar G_k(\xi)$ is smallest for smallest $k$. 
Contributions from other cuts of $k\geq 1$ 
are contributions from instantons. 
Then, the dominant contribution, 
or the contribution from the configuration without the instanton, 
of the integration of the effective potential becomes 
\begin{equation}
 \int_{\text{inside the dominant cut}}\rmd x \e^{-NV_\eff(x)}
  = \sum_n\int_{\mathcal C_0}\rmd x\psi_n^2(x)
  = N . \label{DominantContribution}
\end{equation}
by replacing the summation over n by an integration, 
the contributions from instantons are 
\begin{align}
 \int_{\text{inside sub-dominant cuts}}\rmd x \e^{-NV_\eff(x)} 
 &= N\sum_{k=1}^\infty\sum_n\e^{2\pi i(t+\eta)k} \notag\\
 &\simeq N^2a^2\sum_k\int\rmd \eta \e^{2\pi i(t+\eta)k} \notag\\
 &= -N\sum_k\frac{1}{2\pi ik}\e^{2\pi itk} .
 \label{SubDominantWithoutCorrection}
\end{align}
Here, the summation over $n$ is replaced by the integration 
over $\eta$. 
We also introduced some regularization to suppress 
the contribution from the terms with $\eta\gg 1$, 
because the double scaling limit does not work well 
in the region $\eta\gg 1$ 
and the terms with $\eta\gg 1$ is expected to be nonuniversal. 
Using \eqref{DominantContribution}, 
\eqref{SubDominantWithoutCorrection} 
and $t=i\mu$, 
we obtain 
\begin{equation}
 \delta F = \frac{Z^{\text{(1-inst)}}}{Z^{\text{(0-inst)}}}
  = -\sum_k\frac{1}{2\pi ik}\e^{-2\pi\mu k} .
\end{equation}
But there is further correction to \eqref{SubDominantWithoutCorrection} 
which comes from the fact that we replace the summation over $n$ 
by an integration over $\eta$. 
This correction can be derived via 
the Euler-Mclaurin summation formula. 
Including this correction, 
prefactors of these terms diverge. 
To see this divergence, 
we introduce a regularization. 
When integrating over $\eta$ in \eqref{SubDominantWithoutCorrection}, 
We have introduced some regularization, for example, 
\begin{equation}
 N\sum_k\int\rmd\eta\,\e^{2\pi i(t+\Lambda\eta)k}. 
  \label{IntroducingRegularization}
\end{equation}
We first take $\Impa\Lambda > 0$, 
and after the integration, we take the limit of $\Lambda\to 1$. 
Now, we return the integration over $\eta$ to the summation. 
Then, $\eta$ can be regarded as integer, 
and it can be written as 
\begin{align}
 \int_{\text{inside sub-dominant cuts}}\rmd x \e^{-NV_\eff(x)} 
 &\simeq N\sum_k\sum_{\eta=0}^{\infty}
 \e^{2\pi i(t+\Lambda\eta)k} \notag\\
 &= N\sum_k\frac{1}{1-\e^{2\pi i\Lambda k}}\e^{2\pi itk} . 
 \label{SubDominantWithRegularization}
\end{align}
Then, taking $\Lambda = 1+\epsilon$, we obtain 
\begin{equation}
 \delta F = \frac{Z^{\text{(1-inst)}}}{Z^{\text{(0-inst)}}} 
  = -\sum_k\frac{1}{2\pi i \epsilon k}\e^{-2\pi\mu k} . 
\end{equation}
Therefore, in the limit of $\epsilon\to 0$, 
the prefactors are diverge. 
We discuss this point later. 

Up to now, we have simply considered the correction 
to the free energy from instantons to be 
$\delta F = \frac{Z^{(\text{1-inst})}}{Z^{(\text{0-inst})}}$. 
This is because we have ignored interactions between instantons, 
and we obtain 
\begin{equation}
 \frac{Z^{(n\text{-inst})}}{Z^{(\text{0-inst})}} \simeq 
  {}_NC_n\left(\frac
	  {\int_{\text{inside sub-dominant cuts}} \rmd x
	  \bracket{\det\left(x-\Phi\right)^2}\e^{-NV(x)}}
	  {\int_{\text{inside the dominant cut}} \rmd x
	  \bracket{\det\left(x-\Phi\right)^2}\e^{-NV(x)}}	  
	 \right)^n . 
\end{equation}
Then, the partition function can be written as 
\begin{align}
 Z &\simeq \lim_{N\to\infty} Z^{(\text{0-inst})}
  \left(
   1+\frac{1}{N}\frac{Z^{(\text{1-inst})}}{Z^{(\text{0-inst})}}
  \right)^N \notag\\
  &= \e^{F^{(\text{0-inst})} 
  + \frac{Z^{(\text{1-inst})}}{Z^{(\text{0-inst})}}}. 
\end{align}
Therefore, the contribution 
from instantons to the free energy can be written as 
$\delta F = \frac{Z^{(\text{1-inst})}}{Z^{(\text{0-inst})}}$ 
approximately. 
To be exact, however, there are interactions between instantons, 
and we should include these effect. 
We consider multi-instanton effects in more detail. 

To derive the effective potential of multiple eigenvalues, 
we use a slightly different method from that in \cite{Hanada:2004im}. 
The partition function of the matrix model 
can be written as 
\begin{equation}
 Z = \int\prod_{i=1}^{N}\rmd\lambda\,\Delta^2(\lambda)
  \,\e^{-N\sum_iV(\lambda_i)} . 
\end{equation}
Here, the Vandermonde determinant $\Delta(\lambda)$ 
can be expressed in terms of the orthogonal polynomials as 
\begin{equation}
 \Delta(\lambda) = \det_{mn}P_{n-1}(\lambda_m) . 
\end{equation}
Using this expression we can obtain the effective potential. 
First, we consider the effective potential of one eigenvalue. 
Integrating out the other eigenvalues than 
the $N$-th eigenvalue $x=\lambda_N$, 
we obtain 
\begin{equation}
 Z = (N-1)!\prod_{n=0}^{N-1}h_n\int\rmd x\,
  \sum_{n=0}^{N-1}\psi_n^2(\lambda) . 
\end{equation}
This agrees with \eqref{EffectivePotential_OrthogonalPolynomial}. 
Next, we consider the case 
in which two of eigenvalues are instantons. 
Integrating out $N-2$ eigenvalues, 
excluding $x=\lambda_N$ and $y=\lambda_{N-1}$, 
we obtain 
\begin{equation}
 Z = (N-2)!\prod_{n=0}^{N-1}h_n\int\rmd x\,\rmd y\,\sum_{n\neq m}
  \left(
   \psi_n^2(x)\,\psi_m^2(y)-\psi_n(x)\psi_n(y)\psi_m(x)\psi_m(y)
  \right) . 
\end{equation}
Here, the second term can be dropped because this term 
vanishes due to the orthogonality of 
the orthogonal polynomials.\footnote{
This is a special feature of the Penner model. 
In the case if the $c=0$ matrix model, 
the integration around the saddle point of the instanton 
does not vanish. 
This contributes to the effective potential. 
Then, we cannot simply neglect this term. 
On the other hand, 
integrations inside individual cuts 
vanish separately 
in the case of the Penner model. 
This is because the orthogonal polynomials 
do not depend on the contour. 
The orthogonality condition defined 
on the contour which is along some of cuts 
is satisfied on the contour which is along only one cut. 
Then ,there are no contributions to the effective potential. 
Therefore we can drop this term in the case of the Penner model.}
Then, 
the contribution to the partition function 
from the configuration in which two of eigenvalues are the instanton 
becomes 
\begin{equation}
 Z^{(\text{2-inst})}
  = N!\prod_{n=0}^{N-1}h_n\int_{\text{inside sub-dominant cuts}}\rmd x\,\rmd y
  \sum_{n>m} \psi_n^2(x)\psi_m^2(y) , 
\end{equation}
where we include the factor of ${}_NC_2$, 
which reflects the number of way of specifying 
two isolated eigenvalues. 
we can calculate the contribution from 
the configuration in which $l$ eigenvalues are instanton 
in a similar way. 
It can be written as 
\begin{equation}
 Z^{(l\text{-inst})}
  = N!\prod_{n=0}^{N-1}h_n\int_{\text{inside sub-dominant cuts}}
  \prod_{i=1}^{l}\rmd x_i
  \sum_{n_1>n_2>\cdots>n_l} \prod_{i=1}^l\psi_{n_i}^2(x_i) . 
\end{equation}
By summing up these contributions, 
the partition function can be written as 
\begin{align}
 Z &= Z^{(\text{0-inst})} + Z^{(\text{1-inst})}
 + Z^{(\text{2-inst})}+ \cdots \notag\\
 &= N!\prod_{n=0}^{N-1}h_n\prod_{m=0}^{N-1}
  \left(
   1+\int_{\text{inside sub-dominant cuts}}\rmd x\,
   \psi_m^2(x)
  \right) . 
\end{align}
Then, we calculate this using concrete expressions
of the orthonormal functions. 
We also use the regularization which is introduced 
in \eqref{IntroducingRegularization}. 
Using \eqref{SubDominantWithRegularization}, we obtain 
\begin{equation}
 Z = N!\prod_{n=0}^{N-1}h_n\prod_{m=0}^{N-1}
  \left(1+\sum_{k=1}^{\infty}\e^{2\pi i(t+\Lambda\eta(m))k}\right) . 
\end{equation}
The free energy can be obtained 
by exponentiating this partition function: 
\begin{align}
 \delta F &= \sum_{\eta=0}^{\infty}
 \log\left[1+\sum_{k=1}^\infty\e^{2\pi i(t+\Lambda\eta)k}\right] \notag\\
 &= \sum_{\eta=0}^\infty\log
 \left[\frac{1}{1-\e^{2\pi i(t+\Lambda\eta)}}\right] \notag\\
 &= \sum_{\eta=0}^\infty\sum_{l=0}^\infty
 \frac{1}{l}\e^{2\pi i(t+\Lambda\eta)l} \notag\\
 &= \sum_{l=1}^\infty
 \frac{1}{l\left(1-\e^{2\pi i \Lambda l}\right)}\e^{2\pi itl} . 
\end{align}
Taking $t=i\mu$, we obtain 
\begin{equation}
 \delta F = \sum_{l=1}^\infty
 \frac{1}{l\left(1-\e^{2\pi i \Lambda l}\right)}\e^{-2\pi \mu l} . 
 \label{FreeEnergyWithRegularization}
\end{equation}
Furthermore, we take $\Lambda = 1+\epsilon$. 
Then, $\delta F$ can be written as 
\begin{equation}
 \delta F = \sum_{l=1}^\infty
 -\frac{1}{2\pi i\epsilon l^2}\e^{-2\pi \mu l} . 
\end{equation}
In the limit of $\epsilon\to 0$, 
the prefactors are diverge. 
 
It is known that nonperturbative corrections 
of the $c=1$ noncritical string theory compactified 
on the circle with radius $R$ are given by terms of the form 
$\e^{-2\pi\mu k}$ and $\e^{-2\pi\mu kR}$ with positive integer $k$. 
Our result agrees with these forms. 
Another special feature of $\delta F$ is 
that prefactors of these terms diverge. 
To make the comparison of prefactors with 
those of the nonperturbative effect of 
the $c=1$ noncritical string theory, 
we use not the worldsheet formulation (the Liouville theory) 
but the matrix quantum mechanics formulation. 
The free energy of the $c=1$ noncritical string theory 
with radius $R$ can be obtained via the matrix quantum mechanics 
and expressed as \cite{Alexandrov:2003nn, Alexandrov:2004cg}
\begin{equation}
 F=-\frac{1}{4}\int\frac{\rmd s}{s}\frac{\e^{i\mu s}}
  {\sinh\frac{s}{2}\sinh\frac{s}{2R}}. 
\end{equation}
Taking the contour to be the whole real axis, 
we obtain 
\begin{equation}
  F
  = i\sum_n\frac{\e^{-2\pi n\mu}}
  {4n(-1)^n\sin\left(\frac{\pi n}{R}\right)}
  + i\sum_n\frac{\e^{-2\pi nR\mu}}
  {4n(-1)^n\sin\left(\pi Rn\right)} \label{NPCofMatrixQuantumMechanics}
\end{equation}
These terms come from the residues at the two series of poles 
$s=2\pi i n$ and $s=2\pi inR$. 
There are two types of nonperturbative effects. 
The first term which comes from the poles $s=2\pi i n$ 
doesn't depend on the radius $R$ at leading order of 
the perturbative expansion. 
This term corresponds to the effect due to the D-instanton. 
On the other hand, 
the second term which comes from the poles $s=2\pi inR$ 
depends on the radius $R$ at the leading order of the 
perturbative expansion. 
This term corresponds to the effect due to the D-particle 
which is not localized in the direction compactified on the circle 
with radius $R$. 
For the self-dual radius of $R=1$, 
prefactors of these terms diverge individually. 
However, the whole is regular because 
these divergences cancel each other. 
Therefore the instantons of the Penner model 
correspond to only one of these two types of 
the nonperturbative effect. 
In fact, the expression with regularization 
\eqref{FreeEnergyWithRegularization} 
is quite similar to the individual terms of 
\eqref{NPCofMatrixQuantumMechanics}. 
These two terms cannot be distinguished for $R=1$. 
Someone may think that it is more similar to the first term of 
\eqref{NPCofMatrixQuantumMechanics}, but 
it depends on how to take the regularization. 
We cannot see which 
corresponds to the effect of instantons in the Penner model.

\section{Conclusion}\label{sec:concl}

In this paper, we have studied the correspondence between 
the effects of the instantons in the Penner model and 
the nonperturbative effects in the $c=1$ noncritical string theory 
with self-dual radius. 
We have calculated the contribution from the instantons 
to the free energy of the Penner model. 
In the cases of the model corresponding to the 
$c=0$ noncritical string theory and kazakov series in the 
$c<1$ noncritical string theory, 
the instantons in hermitian one-matrix model 
are studied in recent works. 
The effects of the instantons agree with the 
nonperturbative effects of the noncritical string theory 
in these cases. 
In this paper, 
we have compared the effect of the instantons in the Penner model with 
the nonperturbative effect of the $c=1$ noncritical string theory 
compactified on the circle with self-dual radius of $R=1$. 

In the case of the Penner model, 
especially in the double scaling limit 
corresponding to the $c=1$ noncritical string theory, 
the eigenvalues can no longer regarded as real. 
To make the model well-defined, 
we should introduce an analytic continuation. 
We have extended the definition of the Penner model, 
and defined the partition function 
as a multiple contour integral 
with respect to the $N$ eigenvalues. 
We have also found the suitable choice of the 
both ends of the contour 
that gives the effect of instantons corresponding 
to the nonperturbative effect of the $c=1$ 
noncritical string theory. 
In the case of the $c=0$ matrix model, 
the instanton is defined as the configuration 
with an eigenvalue on the local maximum of the effective potential. 
In the case of the Penner model, 
there are no local maxima of the effective potential.
There are infinite number of cuts. 
In this case, the instanton is defined as an eigenvalue is 
located on the other cut along the contour than 
that corresponding to the perturbative series. 
In this way, 
we have generalized the calculation 
for the instantons to the Penner model. 

Thus, 
we have seen the correspondence between the 
effects of the instanton and nonperturbative effects of 
the $c=1$ noncritical string theory at the leading order. 
The evaluation of the next-to-leading order 
fixes the prefactors of the contribution from the instantons. 
In the Penner model, these prefactors diverge. 
On the other hand, in the $c=1$ noncritical string theory 
with self-dual radius of $R=1$, 
the prefactors of the nonperturbative effect, 
which is calculated in the matrix quantum mechanics, 
do not diverge. 
However, the nonperturbative effects of 
the $c=1$ noncritical string theory 
is constructed from the effects that come from 
the D-instantons and D-particles. 
If we pick up the effects of 
the D-instantons only or D-particles only, 
the prefactor diverges. 
Therefore, the divergence of the prefactor does not 
mean the inconsistency of the correspondence. 
Furthermore, using the expression in which 
we introduce a regularization, 
we can find a similarity between 
our result and nonperturbative corrections 
which is calculated in the matrix quantum mechanics. 
For the self-dual radius of $R=1$, 
we cannot distinguish these effects that come from D-instantons 
and those from D-particles. 
Although we cannot specify which of D-instanton or D-particle 
corresponds to the instantons in the Penner model, 
there should be the effect corresponding to the other. 
And including this, we will obtain the prefactor 
which doesn't diverge. 
This is left for future studies. 

\subsubsection*{Acknowledgements:}

The authors 
are grateful to H.~Kawai 
for valuable advice and discussions. 
We would also like to thank 
M.~Hanada 
for fruitful discussions. 
This work is supported 
in part by the Grant-in-Aid for Scientific Research 
(17$\cdot$1647) 
and 
the Grant-in-Aid for the 21st Century COE 
``Center for Diversity and Universality in Physics'' 
from the Ministry of Education, Culture, Sports, Science 
and Technology (MEXT) of Japan. 
The work of Y.M. is supported in part 
by JSPS Research Fellowships for Young Scientists. 

\appendix

\section{Orthogonal polynomials in the Penner model}\label{app:orth_poly}
The method of orthogonal polynomials is 
a powerful tool to study the hermitian one-matrix model.
In this section, we consider the orthogonal polynomials 
in the large $N$ limit for the Penner model. 

Orthogonal polynomials $P_n(x)=x^n+\mathcal O(x^{n-1})$ 
is defined to obey the following orthogonal condition: 
\begin{equation}
 \left(P_n,P_m\right) = \int\rmd x\ 
  P_n(x)P_m(x)\e^{-NV(x)} = h_n\delta_{nm}. 
  \tag{\ref{orth}}
\end{equation}
The partition function of the hermitian one-matrix model 
is obtained via the recursion relation 
\begin{align}
 xP_n(x) &= X_{nm}P_m(x) = P_{n+1}(x) + s_n P_n(x) + r_n P_{n-1}(x), 
 \tag{\ref{Recursion1}} \\
 P_n'(x) &= \mathcal P_{nm}P_m(x) = \left[NV'(X)\right]_{nm}P_m(x), 
 \tag{\ref{Recursion2}} 
\end{align}
where $r_n = \frac{h_n}{h_{n-1}}$. 
In the case of the Penner model, 
we take the potential $V(x) = \frac{1}{g}\left(x-\log x\right)$, 
and using \eqref{Recursion2}, we obtain 
\begin{subequations}
 \begin{align}
  r_n &= g\xi(g\xi+1), \\
  s_n &= 1+2g\xi+\frac{g}{N},  
 \end{align}
\end{subequations}
where $\xi=\frac{n}{N}$. 
Using the ratio of the orthogonal polynomials 
$\e^{k_n} = \frac{P_n(x)}{P_{n-1}(x)}$, 
\eqref{Recursion1} can be expressed as 
\begin{equation}
 x=\e^{k_{n+1}}+s_n + r_n \e^{-k_n}. \label{eq4k}
\end{equation}
In the large $N$ limit, 
the rescaled index $\frac{n}{N}$ becomes 
a continuous variable $\xi$ and 
$k_n$ becomes continuous function $k(\xi)$. 
Then, we can expand $k$ as 
\begin{subequations}\label{largeNexpand}
 \begin{align}
  k_{n+1} &= k(\xi+\frac{1}{N})= k(\xi) 
  + \frac{1}{N}\partial_\xi k(\xi) + \cdots,
  \label{largeNtaylor}\\
  k_n &= k_n^{(0)} + \frac{1}{N} k_n^{(1)} + \cdots .
 \end{align}
\end{subequations}
In a similar fashion, 
we expand $r_n$ and $s_n$ as
\begin{subequations}
 \begin{align}
  r_n &= r(\xi) =g\xi(g\xi+1), \\
  s_n &= s(\xi) =s_n^{(0)} + \frac{1}{N}s_n^{(1)}, \\
  s_n^{(0)} &= s^{(0)}(\xi) = 1+2g\xi, \\
  s_n^{(1)} &= s^{(1)}(\xi) = g = \frac{1}{2}\partial_\xi s^{(0)}(\xi). 
 \end{align}
\end{subequations}
Substituting these relation to \eqref{eq4k}, 
we obtain 
\begin{subequations}
 \begin{align}
   x&=\e^{k_{n}^{(0)}}+s_n^{(0)} + r_n \e^{-k_n^{(0)}}, \\
  0&=\left(k_n^{(1)}+ \partial_\xi k_n^{(0)}\right)\e^{k_{n}^{(0)}}
  +\frac{1}{2}\partial_\xi s_n^{(0)} - r_n k_n^{(1)}\e^{-k_n^{(0)}}. 
 \end{align}
\end{subequations}
Using these relation, we obtain 
\begin{subequations}\label{sol_of_k}
 \begin{align}
  \e^{k_n^{(0)}} &= \frac{1}{2}
  \left(x- (1+2g\xi) \pm\sqrt{x^2-2(1+2g\xi)x+1}\right), \\
  k_n^{(1)} &= -\frac{1}{4}\partial_\xi
  \log\left[x^2-2(1+2g\xi)x+1\right]. 
 \end{align}
\end{subequations}
The sign of $\pm$ is taken to be $\e^{k_n} \sim x$ at 
$|x|\to\infty$. 
The orthogonal polynomials can be 
expressed as 
\begin{equation}
 P_n(x) = \exp\left[\sum_{m=1}^n k_m\right]. 
\end{equation}
By using the Euler-Maclaurin summation formula 
to convert a summation into an integral, 
it becomes 
\begin{equation}
 P_n(x) = \exp\left[
	       N\int_{0+\frac{1}{2N}}^{\xi+\frac{1}{2N}}
	       \rmd\xi' k(\xi')
	      \right]. 
\end{equation}
The orthonormal functions $\psi_n(x)$ 
is convenient to describe the $n$-dependence of these functions. 
They can be expressed as 
\begin{equation}
 \psi_n(x) = \exp\left[-\frac{N}{2}V(x)+
	       N\int_{0+\frac{1}{2N}}^{\xi+\frac{1}{2N}}
	       \rmd\xi' \left(k(\xi')-\frac{1}{2}\log r(\xi')\right)
	      \right]. 
\end{equation}
Then, substituting \eqref{sol_of_k}, 
we obtain 
\begin{subequations}\label{sol_of_P}
 \begin{align}
  \psi_n(x) 
  &= \left[\frac{\left(x-(1+2g\xi)+\sqrt{x^2-2(1+2g\xi)x+1}\right)^2}
  {4g\xi(g\xi+1)\left(x^2-2(1+2g\xi)x+1\right)}\right]^{\frac14}
  \exp\left[-\frac{N\xi}{2}G_\xi(x)\right] ,  \\
  g\xi G_\xi(x) &= -(1+2g\xi)\log
  \left[\frac{1}{2}\left(x-(1+2g\xi)+\sqrt{x^2-2(1+2g\xi)x+1}\right)\right]
  \notag\\&\qquad
  -\log\left[\frac{1}{2}\left(x+1+\sqrt{x^2-2(1+2g\xi)x+1}\right)\right]
  \notag\\&\qquad
  +\log\left[\frac{1}{2}\left(x-1+\sqrt{x^2-2(1+2g\xi)x+1}\right)\right]
  \notag\\&\qquad
  +\sqrt{x^2-2(1+2g\xi)x+1} -2g\xi.
 \end{align}
\end{subequations}

The orthogonal polynomials $P_n(x)$ are single valued 
on the complex plane of $x$. 
However, its asymptotic form in the large $N$ limit 
has a cut. 
The presence of a cut indicate that the asymptotic form 
of the orthogonal polynomials is not 
single valued. 
This cut is only the consequence of the large $N$ limit. 
We should take the branch which reproduce 
the suitable form of the orthogonal polynomials in the large $N$ limit. 
The cut in the asymptotic form of the orthogonal polynomials 
is identical to that of the resolvent. 
We should take the physical sheet on which the resolvent becomes 
$R(x)\to\frac{1}{x}$ at $x\to\infty$ 
to obtain the suitable branch of the asymptotic behavior 
of the orthogonal polynomials. 
The orthogonal polynomials can be expressed 
in terms of the expectation value of the matrix model as 
\begin{align}
 P_n &= \bracket{\det\left(x-\Phi\right)}_n \notag\\
 &= \exp\left[\bracket{\tr\log\left(x-\Phi\right)}_{c,n}+\cdots\right]. 
\end{align}  
Here, the subscript ``$n$'' indicates that 
the expectation value is taken in the $n\times n$ matrix model, 
and the subscript $c$ indicates connected part. 
Then, in the large $N$ limit, the orthogonal polynomials can be 
expressed in terms of the resolvent $R(x)$ as 
\begin{equation}
 P_n(x) \simeq \exp\left[N\xi\int^x\rmd x'R_\xi(x')\right], \label{orth_and_resolvent}
\end{equation}
where $R_\xi(x)$ is the resolvent with 
the coupling of the matrix model $g$ replaced by $g\xi$. 
In fact, \eqref{sol_of_P} shows that 
$P_n(x)$ behaves as 
\begin{equation}
 P_n(x) \simeq \exp\left[N\int_0^\xi\rmd\xi'k^{(0)}(\xi')\right] 
\end{equation}
at the leading order, and using the relation 
\begin{align}
 \partial_x \int_0^\xi\rmd\xi'k^{(0)}(\xi')
 &= \pm\int_0^\xi\frac{\rmd\xi'}{\sqrt{x^2-2(1+2g\xi')x+1}} \notag\\
 &= \pm\frac{1}{2gx}\left[\sqrt{x^2-2(1+2g\xi)x+1}-|x-1|\right]
 = \xi R_\xi(x), 
\end{align}
we can see the agreement with \eqref{orth_and_resolvent}. 
The sign included in the expression of $k$ should be taken 
to be the same as the resolvent.\footnote{
However, the sign in front of the term of $\frac{1}{2gx}|x-1|$ 
with $x$ in some region 
doesn't agree with the sign of the resolvent. 
This comes from the fact that 
the suitable sign of this term depends on $\xi$. 
If we assume that $k(\xi)$ is continuous function in $\xi$ 
and take the suitable sign for each $\xi$, 
it agrees with the resolvent. 
}
The resolvent becomes $R(x)\sim\frac{1}{x}$ in the limit $x\to\infty$ 
on the physical sheet. 
Then, the orthogonal polynomials behave as 
\begin{equation}
 P_n(x) \sim \exp\left[N\xi\int^x\rmd x'\frac{1}{x'}\right]
  = x^n. 
\end{equation}
It reproduces the suitable behavior of the orthogonal polynomials 
in the limit of $x\to\infty$. 
It also corresponds to that the sign of $\pm$ is 
taken to be $\e^{k_n} \sim x$ at 
$|x|\to\infty$ in \eqref{sol_of_k}. 

On the cut, the orthogonal polynomials can be obtained as 
a linear combination of two branches 
corresponding to, for instance, the $+$ and $-$ in the sign of $\pm$ 
in the \eqref{sol_of_k}. 
It is explained as follows. 
When we consider the orthonormal function 
$\psi_n(x) = \frac{1}{\sqrt{h_n}}P_n(x)\e^{-\frac{N}{2}V(x)}$, 
in the large $N$ limit, it becomes 
\begin{equation}
 \psi_n(x) \sim \exp\left[-\frac{N\xi}{2}\Repa G_\xi(x)\right]
  \sin[\frac{N\xi}{2}\Impa G_\xi(x)+\frac{\pi}{4}]. 
\end{equation}
Since the imaginary part of the $G(x)$ is 
the density of eigenvalues and 
$G_\xi(x)$ is obtained from $G(x)$ by replacing $g$ with $g\xi$, 
the orthonormal functions and orthogonal polynomials 
have the point where $P_n(x) = \psi_n(x) = 0$ on the cut. 
The orthogonal polynomials can be expressed in terms of 
the expectation value of the matrix model as 
\begin{align}
 P_n(x) &= \bracket{\det\left(x-\Phi\right)}_n \notag\\
 &= \bracket{\prod_{i=1}^n(x-\lambda_i)}. 
\end{align}
Then, in the large $N$ limit, 
it becomes $P_n(x)=0$ at the point where an eigenvalue 
is classically located. 
Because the cut of the resolvent comes from 
the support of the eigenvalue distribution, 
the point where $P_n(x)=0$ is on the cut. 
To obtain the point where $P_n(x) = \psi_n(x) = 0$, 
we should take the linear combination of two branches. 
Meanwhile, the oscillatory behavior of $\psi_n^2(x)$ 
can be approximated by the average of $\frac{1}{2}$ 
and we can regard it as a constant inside the cut. 

We can see how to take the linear combination of 
two branches of the asymptotic form of the orthogonal polynomials 
by studying the behavior near the endpoint of the cut. 
In this region, the orthogonal polynomials can be 
approximated by the Airy function. 
In the double scaling limit, 
the orthonormal functions become 
\begin{equation}
 \psi_n \sim \exp\left[-\frac{1}{2}\int\rmd z\sqrt{z^2+4(t+\eta)}\right]. 
\end{equation}
The recursion relation that the orthogonal polynomials satisfy gives 
the differential equation which the orthonormal functions satisfy 
in the double scaling limit. 
The differential equation for the orthonormal functions 
at the leading order of the perturbative expansion 
of the noncritical string is 
\begin{equation}
 \deriv{^2\psi_n}{z^2} = \frac{1}{4}(z^2+4t)\psi_n(z). 
\end{equation}
The both ends of the cut is located on $z=\pm 2i\sqrt t$. 
Introducing the new variable $y$ and $\theta$ defined by 
$z=2i\sqrt t +\e^{i\theta}y$, 
we obtain a differential equation, 
\begin{equation}
  \deriv{^2\psi_n(y)}{y^2} + a\e^{3i\vartheta\frac{\pi}{2}i}y\psi_n(y). 
\end{equation}
Then, for $\theta = \frac{\pi}{6},\,-\frac{\pi}{2},\cdots$, 
we can make use of a standard analysis of the Airy function. 
This implies that the cut is on the line of 
$\theta = \frac{\pi}{6},\,-\frac{\pi}{2},\cdots$. 
It corresponds to the condition that 
the cut is on the line on which 
$G(x)$ is pure imaginary. 
The other condition that 
the cut is embedded on the region 
where $G(x)\leq 0$ 
determines that the cut is on the line of the 
$\theta=\frac{\pi}{6}$. 
The Airy function can be expressed via the integration 
on the complex plane of $k$ as 
\begin{equation}
 \psi(y) = \int\rmd k\ \exp\left[yk+\frac{1}{3}k^3\right]. 
  \label{Airy}
\end{equation}
The asymptotic form of the Airy function can be 
obtained by the method of the steepest descent. 
There are two saddle points in the $k$ plane, 
corresponding to the two branches of the 
asymptotic form of the orthogonal polynomials.  
For $\theta=\frac{\pi}{6}$ and $y>0$, 
we take the contour which picks up the both of 
two saddle points. 
This corresponds to the linear combination 
of the two branches. 
Then, for $y<0$, 
this contour picks up only one of the saddle point. 
So we shouldn't take the linear combination in this region. 
The contour of the integration is determined 
by the choice of the boundary of the contour. 
We should use the same choice of the boundary 
for $\theta=-\frac{\pi}{2}$. 
Since the saddle points are placed at the different point 
from the case of $\theta =\frac{\pi}{6}$, 
this contour picks up only one of the saddle point 
for both $y>0$ and $y<0$. 
In this way, we should take the linear combination 
of two branches only inside the cut. 

The orthonormal functions 
$\psi_n(x)=\frac{1}{\sqrt{h_n}}P_n(x)\e^{-\frac{N}{2}V(x)}$ 
obey the following orthonormality condition: 
\begin{equation}
 \int\rmd x\ \psi_n(x)\psi_m(x) = \delta_{nm}. 
\end{equation}
Here, the integration with respect to $x$ 
can be approximated by the integration inside the cut 
in the large $N$ limit. 
We check \eqref{sol_of_P} satisfies this 
orthonormality condition, especially the normalization. 
First, 
the orthonormal function given by \eqref{sol_of_P} 
becomes 
\begin{equation}
 \psi_n(x) \sim \exp[N\int_0^\xi\rmd\xi'
  \left(k^{(0)}(\xi')-\frac{1}{2}\log r(\xi')\right)] 
\end{equation}
at the leading order of the $\frac1N$ expansion.  
When we take the average of the oscillatory behavior 
of the square of the orthonormal functions $\psi_n^2(x)$, 
it can be regarded as a constant on the cut. 
Examining on the end of the cut $x=x_0(\xi)$, 
we find $k^{(0)}(\xi)-\frac{1}{2}\log r(\xi)=0$. 
Here, the position of the cut depends on the index of 
$n$ of the orthogonal polynomials. 
Then, $n$-dependence of the $\psi_n^2(x)$ becomes 
\begin{equation}
 \Deriv{\psi_n^2(x_0)}{\xi}\sim 
  2N\left(k^{(0)}(\xi)-\frac{1}{2}\log r(\xi)\right)\psi_n^2(x_0)=0. 
\end{equation}
The norm of the orthonormal function $\psi_n(x)$ 
does not depend on $n$ at the leading order of 
the $\frac{1}{N}$ expansion. 

Second, on the correction at the order of $\frac{1}{N}$, 
we have the following expression 
for the orthonormal function $\psi_n(x)$: 
\begin{equation}
 \psi_n^2(x) = \frac{x-(1+2g\xi)+\sqrt{x^2-2(1+2g\xi)x+1}}
  {2\sqrt{x^2-2(1+2g\xi)x+1}\sqrt{g\xi(g\xi+1)}}\e^{\mathcal O(N)}. 
\end{equation}
Here, the $\mathcal O(N)$ contribution can be neglected, 
because it does not depend on $n$ 
as we have seen earlier. 
Using this expression 
and integrating inside the cut, 
we obtain 
\begin{equation}
 \int\rmd x\psi_n^2(x) = 1 .
\end{equation}
Therefore, $\psi_n(x)$ obeys suitable normalization condition 
defined by the integration inside the cut. 

\section{The universality in the Penner model}
Generally, in the double scaling limit, 
the quantities describing the physics of the noncritical string 
do not depend on the details of the potential of the matrix model. 
This universality reflects the fact that 
the physical quantities do not depend on the cut-off. 
Conversely speaking, 
the universal quantity which does not depend on 
the details of the potential 
is physical quantity of the noncritical string theory. 

We can introduce the universality into the Penner model. 
In the case of the Penner model, 
we can add arbitrary polynomials into the potential. 
Generally, we can consider the following potential: 
\begin{equation}
 V(x) = \frac{1}{g}\left(\sum_{k=1}^l\frac{g_k}{k}x^k
		   -\log x\right). 
\end{equation}
When we consider the Penner model as the model 
corresponding to the $c=1$ noncritical string theory, 
we restrict ourselves to the critical point 
on which two ends of the cut merge. 
On such critical point, 
$r_n$ in \eqref{Recursion1} takes the value $r_c=0$. 
In the double scaling limit, that is the $N\to\infty$ with 
\begin{align}
 g&=g_c(1-a^2 t), &
 x&=x_c+az, &
 \xi & = 1-a^2\eta, &
 N&= a^{-2},  
\end{align}
$r_n$ and $s_n$ defined in \eqref{Recursion1} becomes 
\begin{subequations}
 \begin{align}
  r_n &= -a^2\alpha\ (t+\eta) + \mathcal O(a^4), \\
  s_n &= x_c + a^2\beta\ (t+\eta) + \mathcal O(a^4).
 \end{align}
\end{subequations}
The resolvent can be described in terms of 
the orthogonal polynomials as 
\begin{align}
 R(x) &= \int_0^1\rmd\xi\oint\frac{\rmd\omega}{\omega}
 \frac{1}{\omega+s(\xi)+r(\xi)\,\omega^{-1}}\notag\\
 &\simeq \int\frac{\rmd\eta}{\sqrt{z^2+4\alpha\ (t+\eta)}} \notag\\
 &= \frac{1}{2\alpha}\sqrt{z^2+4\alpha\ (t+\eta)}.
\end{align}
Then, $\bar G_{k}(\xi)$ in this case becomes
\begin{align}
 \bar G_{k+1}(\xi)-\bar G_{k}(\xi) &= \int\rmd z
 \frac{1}{\alpha}\sqrt{z^2+4\alpha\ (t+\eta)} \notag\\
 &= -2\pi i(t+\eta). 
\end{align}
It agrees with the result for the usual Penner model \eqref{G-G}. 
We can do the rest of the calculation in the same way as 
the usual Penner model. 
Then, we obtain the same result for the contribution 
from the instantons.

\end{document}